\documentclass[10pt, twocolumn, comsoc]{IEEEtran}

\def\argmax{\mathop{\rm arg\,max}}

\usepackage{graphicx,epsfig}
\usepackage[noadjust]{cite}
\usepackage{amsfonts,helvet}
\usepackage{threeparttable}
\usepackage{amsthm}
\usepackage{amsmath}
\usepackage{siunitx}
\usepackage{amssymb}
\usepackage{dsfont}
\usepackage{color}
\usepackage[linesnumbered,ruled]{algorithm2e}
\usepackage{enumerate}
\usepackage{cancel}
\usepackage{comment}
\usepackage{placeins}

\usepackage[colorlinks=true, linkcolor=blue]{hyperref}

\newcommand{\fig}[1]{Fig.\ \ref{#1}}

\newtheorem{remark}{Remark}

\SetKwInput{KwInput}{Initialize}
\SetKwInput{KwOutput}{Output}

\usepackage{eucal}

\begin{document}
\bstctlcite{IEEEexample:BSTcontrol}

\title{Soft-then-Hard: Joint Link Association and Precoding Optimization for LEO Satellite Networks}

\author{Hyeongtak~Yun, Wonjae~Shin, and Jeonghun~Park

\thanks{This work was supported in part by the Institute of Information \& communications Technology Planning \& Evaluation (IITP) grant funded by the Korea government (MSIT) (No. RS-2024-00397216, Development of the Upper-mid Band Extreme massive MIMO (E-MIMO)), and in part by the Samsung Research Funding \& Incubation Center of Samsung Electronics under Project Number SRFC-IT2402-06.

Hyeongtak Yun and Jeonghun Park are with the School of Electrical and Electronic Engineering, Yonsei University, Seoul 03722, South Korea (e-mail: {\texttt{yht3114@yonsei.ac.kr; jhpark@yonsei.ac.kr}}). Wonjae Shin is with the School of Electrical Engineering, Korea University, Seoul 02841, South Korea (e-mail: {\texttt{wjshin@korea.ac.kr}}).}}

\maketitle \setcounter{page}{1}

\begin{abstract}
Spatial multiplexing in dense low Earth orbit (LEO) satellite downlinks is hindered by two bottlenecks. First, line-of-sight (LoS)-dominant propagation from a high orbital altitude places the users in a narrow angular region, which limits the spatial resolution achievable by a single satellite. Second, the visible satellites cause severe inter-satellite interference. Both effects call for joint link association and precoding, because the association decides which users are served together and which streams interfere. However, conventional block coordinate descent with hard association updates suffers from premature turning-off, where an inactive link loses precoder power and cannot be reactivated in subsequent iterations. To overcome this limitation, we propose a soft-then-hard update scheme that relaxes the association into continuous weights via iterative scaling, keeping candidate links active. Annealing progressively concentrates the association weights as the precoders are updated, after which a single hard matching determines the serving links. We further extend this framework to the LEO space-time channel, where distinct Doppler shifts provide user separability beyond angular geometry. Numerical results show that, over the space-time channel, the proposed design outperforms the baselines while effectively exploiting this separability.
\end{abstract}


\section{Introduction}

Low Earth orbit (LEO) satellite communications have emerged as a promising means of extending connectivity to areas underserved by terrestrial mobile networks. Compared with geostationary (GEO) satellites, LEO satellites operate at much lower altitudes and thereby reduce propagation loss and latency considerably, at the cost of a smaller per-satellite footprint that makes a constellation of many satellites necessary for continuous service. Commercial systems such as Starlink and OneWeb have been deploying constellations comprising hundreds to thousands of satellites and now offer broadband services over wide areas, and Amazon Leo (formerly Project Kuiper) is deploying its own constellation~\cite{kodheli:comst:21}. With wide-area connectivity increasingly in place, the design emphasis is shifting toward how much traffic can be accommodated under the limited spectrum and onboard power, i.e., toward high throughput.

A direct way to increase throughput is to serve more users on the same time and frequency resources, and multiuser multiple-input multiple-output (MU-MIMO) precoding enables such spatial multiplexing by separating users in the spatial domain through the onboard antenna array~\cite{yun:twc:26, zheng:twc:12, vazquez:wcm:16}. The geometry of the LEO downlink, however, is unfavorable to spatial multiplexing in two respects. First, the channel is dominated by the line-of-sight (LoS) path with little local scattering, so users can be distinguished only through their angular differences as seen from the array. Because the orbital altitude greatly exceeds the cell radius, unlike in typical terrestrial cellular networks, these angular differences are small relative to the resolution of the onboard array. The user channel vectors are then highly correlated, limiting the number of users that a single satellite can effectively multiplex to well below that suggested by the array size~\cite{you:jsac:20, yun:twc:26}. Second, as constellations densify under full frequency reuse (FFR), multiple visible satellites cause severe inter-satellite interference. Unlike in terrestrial networks, the vast distance to orbit means that interfering satellites are not much farther than the serving satellite, so interference arrives almost as strong as the desired signal. Deploying more satellites may therefore even degrade coverage instead of improving it~\cite{park:twc:23, kim:arxiv:25}.

Because of these structural limitations, simultaneous MU-MIMO transmission from multiple satellites is more difficult than in terrestrial cellular networks, and the attainable spatial multiplexing gain is reduced. Overcoming them requires careful attention to the interference topology among satellite--user links. This topology is shaped by the link association, which specifies both the serving satellite of each user and the set of users that each satellite has to separate by precoding. Link association and multi-satellite precoding must therefore be designed jointly to manage intra- and inter-satellite interference and enhance the sum spectral efficiency. 


However, the resulting joint association and precoding problem is challenging because it couples discrete association decisions with nonconvex precoder optimization.
A widely adopted approach to resolve this difficulty is block coordinate descent (BCD), which alternately updates the association and the precoders until convergence~\cite{shen:tsp:18:part1, ammar:twc:22a}. When the association subproblem is solved using an exact hard matching at each iteration, however, a failure mode known as premature turning-off arises~\cite{shen:tsp:18:part2, shen:ton:19:matrixfp}.
Specifically, once a link becomes inactive, the subsequent precoder update assigns it zero power. Its association score then vanishes, leaving no incentive to reactivate the link in later iterations. As a result, potentially useful links may be discarded before their precoders are sufficiently optimized. Since these links cannot be recovered, the links carrying nonzero power remain a subset of the initial association, making the final performance strongly dependent on the initialization.


Compounding this difficulty, the dense geometry also limits what the association itself can accomplish. Although the candidate satellites may be separated by hundreds of kilometers in orbit, the slant range from each of them to the cell far exceeds the spacing between users, so every candidate satellite observes the users clustered within a narrow angular sector~\cite{you:jsac:20, yun:twc:26}. Reassigning a user therefore moves that user to a different satellite without improving the angular geometry that the new satellite faces. As long as the design relies on the spatial domain alone, even a joint design cannot supply the spatial degrees of freedom that the geometry does not provide.

In this paper, to address these algorithmic and physical limitations, we develop a soft-then-hard association and precoding design framework that also exploits the Doppler differences among satellite--user links. During the iterative optimization process, the binary association is relaxed to continuous weights, and the resulting association problem is formulated as an entropic optimal transport problem that can be efficiently solved by the Sinkhorn algorithm~\cite{sinkhorn-knopp:1967, cuturi:2013}. Owing to the entropic regularization, every candidate link stays active with a nonzero weight, so potentially useful links are not discarded before their precoders have been sufficiently optimized. In addition, we stack temporal snapshots into a space-time channel~\cite{yim:twc:26, yun:twc:26} and thus augment the highly correlated spatial channels with the distinct Doppler shifts of the links, which are given by the satellite--user relative velocities projected onto the link directions. The added temporal dimension does more than strengthen individual links: it also enables users with similar spatial directions to be distinguished by their Doppler differences. How much the space-time channel helps therefore depends on whether the link association selects users that can exploit this additional separation. Since candidate links remain active throughout the soft phase, the proposed framework can evaluate such links together with their precoders before the association is fixed. In this way, soft association preserves links that can benefit from Doppler separation, while the space-time channel provides additional separability for these links, jointly addressing the algorithmic and physical limitations.

\subsection{Related Works}
\label{subsec:related_works}

One way to mitigate interference in LEO satellite downlinks is to coordinate resource usage to avoid mutual interference in the same time-frequency slots. In~\cite{lin:tcom:23}, a beam-hopping scheme combining load balancing and interference avoidance was proposed for multi-satellite non-geostationary orbit (NGSO) systems, and it was extended in~\cite{lin:twc:24} to jointly determine the cell--satellite association and the multi-satellite beam-hopping schedule. While these beam-hopping methods can alleviate interference by restricting the simultaneous activation of beams that cause strong interference, they also limit the opportunities to simultaneously reuse the same resources for multiple transmissions.

As advances in radio frequency and digital hardware technologies have made it feasible to equip satellites with phased-array antennas, beamforming and precoding techniques have been studied as a complementary approach to directly suppress interference while maintaining high frequency reuse. For instance, \cite{you:jsac:20, li:tcom:22} designed downlink precoders based on statistical channel state information (CSI) to suppress multiuser interference in LEO massive MIMO systems under FFR, where~\cite{you:jsac:20} additionally grouped users by their space angles and~\cite{li:tcom:22} maximized the ergodic sum rate. This was extended in~\cite{you:twc:22:hybrid-leo} to hybrid analog/digital precoding to jointly consider implementation complexity and energy efficiency. Furthermore, \cite{wang:twc:24} jointly optimized beam activation and beamforming in beam-hopping LEO systems, considering both resource-domain interference avoidance and spatial-domain interference suppression.

In dense LEO constellations, where multiple satellites are simultaneously visible from the same region, the scope of interference management has expanded from single-satellite precoding to multi-satellite coordination. The performance of LEO networks using dynamic coordinated beamforming was analyzed in~\cite{kim:twc:24}, demonstrating coverage and spectral efficiency gains when coordinating multiple satellites. In~\cite{xu:tcom:24:drsma}, rate-splitting multiple access was implemented in a distributed manner at the GEO and LEO satellites of a multilayer system to manage the interference between the two layers with statistical CSI. In~\cite{zhang:twc:24:multisat}, hybrid beamforming and user scheduling were jointly designed in multi-satellite networks to consider both satellite--user connection and interference suppression.

However, the multiuser transmission capability of these spatial-domain techniques is limited by how well the spatial channels of simultaneously active users are separated. Especially in LoS-dominant LEO downlinks, a small terrestrial service area is viewed from a high orbital altitude, so the spatial directions of the users are similar and their channels are strongly correlated~\cite{you:jsac:20, yun:twc:26}. Therefore, it is difficult to fundamentally resolve these geometric limitations through spatial processing alone. To overcome these spatial limitations, \cite{yim:twc:26, yun:twc:26} proposed space-time beamforming that exploits Doppler shifts as an additional signaling dimension, without determining which satellite serves which user.

Scheduling determines the interference topology, whereas beamforming controls the desired signal power and the interference leakage. This coupling has motivated extensive research on joint scheduling and beamforming~\cite{yu:twc:13:scheduling, hassan:ett:14, gesbert:jsac:10, bjornson:fnt:13}. The resulting problem is a nonconvex mixed-integer program and is NP-hard in general~\cite{liu:tsp:14:ofdma}. In terrestrial multicell networks, it has typically been tackled by BCD, which alternates between the discrete schedule and the continuous beamformers; for a given schedule, the beamformers are updated via the weighted minimum mean square error (WMMSE) algorithm~\cite{shi:tsp:11} or fractional programming (FP), and the schedule is then updated via per-beam greedy selection or weighted Hungarian matching between the users and the current beamformers~\cite{yu:twc:13:scheduling, khan:twc:20}.

However, hard scheduling updates within BCD that zero inactive beamformers cause premature turning-off~\cite{shen:tsp:18:part2, shen:ton:19:matrixfp}, where candidate links are discarded before being sufficiently evaluated. The WMMSE algorithm is vulnerable to this problem since a zero beamformer is a fixed point of its update~\cite{shen:ton:19:matrixfp, khan:twc:20}, and the sparse beamforming design in~\cite{dai:access:14} further removes links with vanishing beamformers from the candidate set to reduce complexity. In~\cite{khan:twc:20}, FP is combined with Hungarian matching so that all users, including unscheduled ones, are evaluated at every iteration, which alleviates this problem. Nevertheless, each candidate is evaluated with a beamformer adapted to a previously scheduled user, and the final performance remains sensitive to initialization.

For multi-satellite LEO downlinks, the joint design of association and precoding has received little attention. In~\cite{zhang:twc:24:multisat}, links are added greedily according to the increase in the sum spectral efficiency without exploring all combinations of candidate links, so that links discarded at an early stage are never reconsidered, which is essentially premature turning-off. In addition, since the users are spatially clustered in LEO downlinks, there is little room to separate them in the spatial domain, and thus optimizing the association over these candidate links alone provides only limited gains.

To overcome both the fundamental spatial limitation of LEO downlinks and premature turning-off, we propose soft association with annealing under space-time precoding. Deterministic annealing and entropic optimal transport have been applied to association problems, e.g., beam placement in NGSO satellite systems and device association in cloud radio access networks~\cite{torkzaban2023capacitated, paschos2018computational}. However, since these works do not optimize the association jointly with multiuser precoding, the obtained association tends to be suboptimal for the joint problem.

\subsection{Contributions}
\label{subsec:contributions}

The main contributions of this paper are summarized as follows:
\begin{itemize}
\item We develop a joint link association and precoding framework for multi-satellite LEO downlinks under FFR. FP yields closed-form precoder updates, and the derived association score accounts for both received and caused interference. We also identify how zeroing inactive precoders causes both their power and scores to remain zero under subsequent block updates.
\item We propose a soft-then-hard association update that mitigates premature turning-off. Entropic optimal transport, solved by the Sinkhorn algorithm, keeps every association weight positive while the candidate precoders are updated, and annealing progressively concentrates the weights before the binary serving links are determined once under the serving caps.
\item We extend the framework to space-time precoding so that association decisions account for Doppler separation as well as angular separation. This enables the joint design to select user combinations that spatial precoding alone separates poorly.
\item Numerical results show that the proposed scheme outperforms the considered baselines in the space-time setting. At the nominal serving capacity, space-time processing improves the proposed scheme, whereas its repetition cost reduces the performance of the FP baselines at high power. Moreover, determining the association from the slowly varying channel geometry while only the precoders track instantaneous fading retains an advantage over the baselines in the considered LoS-dominant setting.
\end{itemize}

\textit{Notation:} Scalars, vectors, and matrices are denoted by $a$, $\mathbf{a}$, and $\mathbf{A}$, respectively. $(\cdot)^{\mathsf{T}}$, $(\cdot)^{\mathsf{H}}$, $(\cdot)^{\ast}$, and $(\cdot)^{-1}$ denote transpose, Hermitian transpose, conjugate, and inverse. $\Re\{\cdot\}$, $\mathbb{E}[\cdot]$, $\vert{}\cdot\vert{}$, $\Vert{}\cdot\Vert{}$, and $\otimes$ denote real part, expectation, absolute value, Euclidean norm, and Kronecker product. $\mathbf{I}_{M}$ denotes the $M \times M$ identity matrix, and $\mathbf{0}_{M}$ denotes the $M \times 1$ zero vector. $\mathbf{A} \succeq \mathbf{0}$ means that $\mathbf{A}$ is positive semidefinite. $\mathcal{CN}(\mu,\sigma^2)$ denotes the circularly symmetric complex Gaussian distribution with mean $\mu$ and variance $\sigma^2$.

\section{System Model}
\label{sec:system}
\subsection{Network Model}
\label{subsec:network_model}
We consider multi-satellite LEO downlinks serving a set of ground users $\mathcal{K}=\{1,\dots,K\}$. Each satellite is equipped with a uniform planar array (UPA) of $M=M_xM_y$ elements with half-wavelength spacing along the $x$ and $y$ axes. Each user is equipped with a single antenna and is located within a circular cell of radius $R$ centered at a ground reference point.

\begin{figure}[!t]
\centering
\includegraphics[width=0.9\columnwidth]{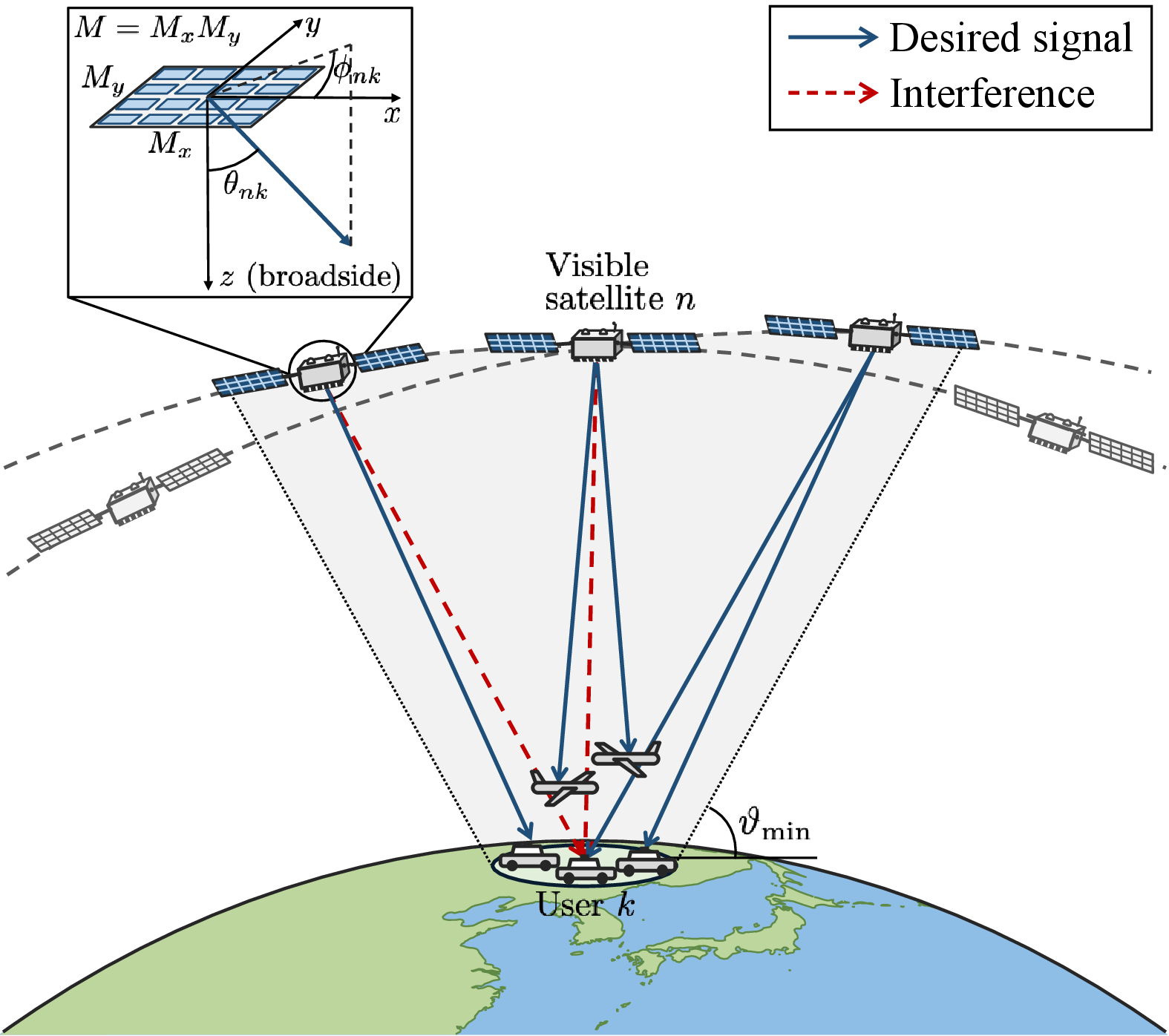}
\caption{System model of multi-satellite LEO downlink communications. Blue solid lines denote the desired signals to the users, and red dashed lines denote the interference to user $k$.}
\label{fig:system}
\end{figure}

Since path loss and atmospheric attenuation increase rapidly at low elevation angles, we consider only satellites whose elevation angle at the cell center is at least a minimum threshold $\vartheta_{\min}$ for service. We denote the set of such visible satellites by $\mathcal{N}=\{1,\dots,N\}$, so that $\vartheta_n\geq\vartheta_{\min}$ for all $n\in\mathcal{N}$, where $\vartheta_n$ is the elevation angle of satellite $n$ at the cell center. \fig{fig:system} illustrates the considered system model.

A ground gateway coordinates the link association and precoding decisions and is assumed to have CSI for all satellite--user links. Each user obtains its position and velocity from a global navigation satellite system (GNSS) and reports them to the gateway, which also tracks the satellite ephemeris. Hence, the channel parameters of each link, namely the path gain, spatial frequency, and Doppler shift, can be obtained at the gateway with minimal signaling overhead~\cite{3gpp_tr38821}.

\subsection{Channel Model}
\label{subsec:generic_channel}
Following the LEO multipath channel models in~\cite{you:jsac:20, li:tcom:22}, the downlink channel between satellite $n$ and user $k$ at carrier frequency $f_{\mathrm{c}}$ is modeled as
\begin{equation}
\mathbf{g}_{nk}(t) = \sqrt{M} \sum_{p=0}^{N_{nk}^{\mathrm{p}}-1} \alpha_{nk,p} e^{j2\pi(\nu_{nk,p}t - f_{\mathrm{c}}\tau_{nk,p})} \mathbf{a}_{nk,p},
\end{equation}
where $N_{nk}^{\mathrm{p}}$ is the number of propagation paths, and $\alpha_{nk,p}$, $\tau_{nk,p}$, $\nu_{nk,p}$, and $\mathbf{a}_{nk,p}$ denote the complex gain, propagation delay, Doppler shift, and array response vector of the $p$-th path, respectively.

The array response for path $p$ is given by
\begin{equation}
\mathbf{a}_{nk,p} = \mathbf{a}(u^y_{nk,p}; M_y) \otimes \mathbf{a}(u^x_{nk,p}; M_x),
\end{equation}
where $\mathbf{a}(u; M) = \frac{1}{\sqrt{M}} [1, e^{j2\pi u}, \dots, e^{j2\pi(M-1)u}]^{\mathsf{T}}$ represents the normalized array steering vector for each axis. The spatial frequencies for the two axes are defined as
\begin{equation}
\begin{aligned}
u^x_{nk,p} &= \frac{1}{2}\sin\theta_{nk,p}\cos\phi_{nk,p}, \\
u^y_{nk,p} &= \frac{1}{2}\sin\theta_{nk,p}\sin\phi_{nk,p},
\end{aligned}
\end{equation}
where $\theta_{nk,p}$ is the zenith angle measured from the UPA broadside and $\phi_{nk,p}$ is the azimuth angle measured from the $x$-axis on the array plane.

Since the local scattering region around the user is small compared with the distance to the satellite, all propagation paths within the same link are observed from nearly the same direction at the satellite array. Accordingly, we assume a common array response for all paths, i.e., $\mathbf{a}_{nk,p} = \mathbf{a}_{nk}$.

Moreover, since high satellite velocity dominates the channel dynamics, all propagation paths share a common dominant Doppler shift $\nu_{nk}$. The time-varying channel thus naturally factorizes as
\begin{equation}
\mathbf{g}_{nk}(t) = \sqrt{M} \beta_{nk}(t) e^{j2\pi\nu_{nk}t} \mathbf{a}_{nk},
\label{eq:reduced}
\end{equation}
where $\beta_{nk}(t) = \sum_{p=0}^{N_{nk}^{\mathrm{p}}-1} \alpha_{nk,p} e^{j2\pi(\Delta\nu_{nk,p}t - f_{\mathrm{c}}\tau_{nk,p})}$ accounts for propagation delays and residual Doppler shifts $\Delta\nu_{nk,p} = \nu_{nk,p} - \nu_{nk}$. 
For spatial precoding, evaluating the channel at reference time $t=0$ yields the static model $\mathbf{h}_{nk}$ in~\eqref{eq:channel}, where $\beta_{nk} = \beta_{nk}(0)$. Section~\ref{sec:stab} subsequently restores the temporal phase progression $e^{j2\pi\nu_{nk}t}$ to separate users through Doppler differences across links.

Since the LoS path dominates LEO downlinks~\cite{3gpp_tr38811}, we model the composite gain as Rician fading with a complex Gaussian diffuse term,
\begin{equation}
\beta_{nk} = \sqrt{\eta_{nk}} \left( \sqrt{\frac{\kappa_{nk}}{1+\kappa_{nk}}} e^{j\psi_{nk}} + \sqrt{\frac{1}{1+\kappa_{nk}}} w_{nk} \right),
\label{eq:beta_gen}
\end{equation}
where $\kappa_{nk}$ is the Rician factor, $\psi_{nk}$ is the phase of the LoS component, and $w_{nk} \sim \mathcal{CN}(0,1)$ represents the aggregate diffuse component. The average channel power $\eta_{nk}$ is given by
\begin{equation}
\eta_{nk} = G_{\mathrm{r}} \left( \frac{c}{4\pi f_{\mathrm{c}} d_{nk}} \right)^{2},
\end{equation}
where $d_{nk}$ is the distance between satellite $n$ and user $k$, $G_{\mathrm{r}}$ is the receive antenna gain of the user, and $c$ is the speed of light. Consequently, the instantaneous spatial channel is modeled as
\begin{equation}
\mathbf{h}_{nk} = \mathbf{g}_{nk}(0) = \sqrt{M} \beta_{nk} \mathbf{a}_{nk} \in \mathbb{C}^M.
\label{eq:channel}
\end{equation}

\subsection{Signal Model}
\label{subsec:signal_model}
The visible satellite set $\mathcal{N}$ specifies the candidate serving satellites, while the active satellite--user links remain to be determined. We represent the link association by a binary variable $x_{nk}\in\{0,1\}$, where $x_{nk}=1$ if satellite $n$ serves user $k$ and $x_{nk}=0$ otherwise. The association variables are collected into the matrix $\mathbf{X}=[x_{nk}]\in\{0,1\}^{N\times K}$. We assume that each user is served by at most one satellite.

Under FFR, all visible satellites transmit over the same time-frequency resource. Satellite $n$ applies a precoder $\mathbf{f}_{nk}\in\mathbb{C}^{M}$ to the data symbol $s_k\sim\mathcal{CN}(0,1)$ for user $k$. We collect the precoders of satellite $n$ as $\mathbf{F}_n=[\mathbf{f}_{n1},\dots,\mathbf{f}_{nK}]\in\mathbb{C}^{M\times K}$ and denote their collection by $\mathbf{F}=\{\mathbf{F}_n\}_{n=1}^{N}$. The transmit signal of satellite $n$ is then given by
\begin{equation}
\mathbf{x}_n = \sum_{k\in\mathcal{K}} x_{nk}\,\mathbf{f}_{nk}\,s_k
\in\mathbb{C}^{M}.
\label{eq:tx}
\end{equation}
Since the data symbols are mutually independent and have unit power, the average transmit power of satellite $n$ is
$\mathbb{E}[\|\mathbf{x}_n\|^2]
=\sum_{k\in\mathcal{K}}x_{nk}\|\mathbf{f}_{nk}\|^2$,
which is limited by the transmit power budget $P_n$. We set $\mathbf{f}_{nk} = \mathbf{0}_{M}$ whenever $x_{nk} = 0$, so that no precoding vector is assigned to an inactive satellite--user link.

The received signal of user $k$ is
\begin{equation}
y_k =
\underbrace{\sum_{n\in\mathcal{N}} x_{nk}\,
\mathbf{h}_{nk}^{\mathsf{H}}\mathbf{f}_{nk}s_k}_{\text{desired signal}}
+
\underbrace{\sum_{n\in\mathcal{N}}\sum_{k'\neq k} x_{nk'}\,
\mathbf{h}_{nk}^{\mathsf{H}}\mathbf{f}_{nk'}s_{k'}}_{\text{interference}}
+z_k,
\label{eq:rx}
\end{equation}
where $z_k\sim\mathcal{CN}(0,\sigma^2)$ denotes the additive white Gaussian noise (AWGN). The first term is the signal from the serving satellite, and the second is the interference from the streams intended for the other users over all visible satellites.

The signal-to-interference-plus-noise ratio (SINR) of user $k$ when served by satellite $n$ is
\begin{equation}
\mathrm{SINR}_{nk}(\mathbf{X},\mathbf{F}) =
\frac{\big|\mathbf{h}_{nk}^{\mathsf{H}}\mathbf{f}_{nk}\big|^2}
{\sigma^2 + \sum_{n'\in\mathcal{N}}\sum_{k'\neq k}
x_{n'k'}\,\big|\mathbf{h}_{n'k}^{\mathsf{H}}\mathbf{f}_{n'k'}\big|^2}.
\label{eq:sinr}
\end{equation}

Using natural logarithms, the achievable sum spectral efficiency of the network is then
\begin{equation}
R_{\Sigma}(\mathbf{X},\mathbf{F}) =
\sum_{n\in\mathcal{N}}\sum_{k\in\mathcal{K}} x_{nk}\,
\log\!\big(1+\mathrm{SINR}_{nk}(\mathbf{X},\mathbf{F})\big).
\label{eq:sumse}
\end{equation}

\subsection{Problem Formulation}
\label{subsec:formulation}

The association $\mathbf X$ determines the active links, the users served by each satellite, and the interfering streams received by each user. It therefore specifies the interference environment for which $\mathbf F$ is designed. Conversely, the benefit of activating a link depends on $\mathbf F$, which controls interference suppression and the allocation of each satellite's power budget $P_n$ among its active links. We therefore optimize the association and precoders jointly to maximize the sum spectral efficiency.
Accordingly, we formulate the joint design as
\begin{subequations}\label{eq:problem}
\begin{align}
(\mathcal{P}_0)\quad \max_{\mathbf{X},\mathbf{F}}\quad
& R_{\Sigma}(\mathbf{X},\mathbf{F}) \label{eq:obj}\\
\mathrm{s.t.}\quad
& \sum_{n\in\mathcal{N}} x_{nk} \le 1, && \forall k\in\mathcal{K}, \label{eq:c_assoc}\\
& \sum_{k\in\mathcal{K}} x_{nk} \le K_n, && \forall n\in\mathcal{N}, \label{eq:c_cap}\\
& \sum_{k\in\mathcal{K}} x_{nk}\|\mathbf{f}_{nk}\|^2 \le P_n, && \forall n\in\mathcal{N}, \label{eq:c_power}\\
& x_{nk}\in\{0,1\}, && \forall n\in\mathcal{N},\ k\in\mathcal{K}. \label{eq:c_binary}
\end{align}
\end{subequations}

Constraint \eqref{eq:c_assoc} ensures that each user is associated with at most one satellite, while allowing the optimization to select a subset of users for service. Constraint \eqref{eq:c_cap} limits satellite $n$ to simultaneously serving at most $K_n$ users over the same resource. Constraint \eqref{eq:c_power} enforces the power budget of each satellite, and \eqref{eq:c_binary} restricts the link association to binary values.
The difficulty of $(\mathcal{P}_0)$ comes from both the discrete association and the nonconvex objective. The association variables select the active rate terms and the interfering streams, and they are coupled with the precoders through the power constraint. The precoders enter both the numerator and the denominator of every SINR, so the objective is a sum of logarithms of ratios. Even for a fixed association, maximizing such a sum of ratios is NP-hard, so a globally optimal solution of $(\mathcal{P}_0)$ is intractable in general.

\begin{remark}[Effective spatial multiplexing capability] \normalfont \label{rem:cap}
The serving capacity $K_n<M$ reflects the effective spatial multiplexing capability of the LEO downlink rather than the array size. Because the satellite views the ground cell from a high altitude through LoS links, the users lie within a narrow angular sector and their steering vectors are strongly correlated. The multiuser channel therefore becomes ill-conditioned as more users are multiplexed, and the antenna count $M$ does not translate into as many spatial streams~\cite{you:jsac:20}. The cap therefore limits the number of simultaneously served users and preserves spatial degrees of freedom for suppressing residual interference.
\end{remark}
\section{Problem Reformulation}
\label{sec:reformulation}
We solve $(\mathcal{P}_0)$ by BCD. Even with the association fixed, the precoder subproblem is a nonconvex sum-of-ratios problem without a tractable update. We therefore first introduce auxiliary variables through FP so that each block has a closed-form update, and then update the auxiliary variables, the precoders, and the association alternately.

\subsection{Fractional Programming Reformulation}
\label{subsec:problem_reformulation}
With the association fixed, the objective is still a sum of ratios whose numerators and denominators are coupled through the precoders. FP handles this form with two transforms~\cite{shen:tsp:18:part1,shen:tsp:18:part2}. The Lagrangian dual transform followed by the quadratic transform yields an equivalent reformulation with auxiliary variables, which is concave in the precoders when the association and the auxiliary variables are fixed, so that the precoders admit a closed-form update.
The Lagrangian dual transform~\cite{shen:tsp:18:part2} introduces an auxiliary variable $\gamma_{nk}\geq0$ for each link, collected in $\boldsymbol{\gamma}=\{\gamma_{nk}\}$, and turns the sum spectral efficiency into
\begin{equation}
\label{eq:fr_ldt}
\begin{aligned}
f_{\mathrm r}(\mathbf X,\mathbf F,\boldsymbol{\gamma}) = \sum_{n\in\mathcal N}\sum_{k\in\mathcal K}x_{nk}
\Bigg[
&\log(1+\gamma_{nk})-\gamma_{nk} \\
&+ \frac{(1+\gamma_{nk})S_{nk}(\mathbf F)}{S_{nk}(\mathbf F)+I_{nk}(\mathbf X,\mathbf F)}
\Bigg],
\end{aligned}
\end{equation}
where
\begin{equation}
S_{nk}(\mathbf F) = \left|\mathbf h_{nk}^{\mathsf H}\mathbf f_{nk}\right|^2
\label{eq:S_nk}
\end{equation}
denotes the desired signal power of link $(n,k)$, and
\begin{equation}
I_{nk}(\mathbf X,\mathbf F) = \sigma^2 + \sum_{n'\in\mathcal N}\sum_{k'\neq k} x_{n'k'} \left|\mathbf h_{n'k}^{\mathsf H}\mathbf f_{n'k'}\right|^2
\label{eq:I_nk}
\end{equation}
denotes the corresponding interference-plus-noise power.
For fixed $(\mathbf X,\mathbf F)$, $f_{\mathrm r}$ is concave in each $\gamma_{nk}$, and its maximizer is
\begin{equation}
\gamma_{nk}^{\star} = \frac{S_{nk}(\mathbf F)}{I_{nk}(\mathbf X,\mathbf F)} = \mathrm{SINR}_{nk}(\mathbf X,\mathbf F).
\label{eq:gamma_opt}
\end{equation}
Substituting $\gamma_{nk}^{\star}$ back recovers the sum spectral efficiency, so the reformulation is equivalent to $(\mathcal{P}_0)$.

The precoders appear in both the numerators and denominators of the coupled ratios in $f_{\mathrm r}$, making the objective nonconcave. FP problems with multiple ratios are NP-hard in general~\cite{shen:tsp:18:part1}. The quadratic transform~\cite{shen:tsp:18:part1} decouples it with a complex auxiliary variable $\zeta_{nk}$ for each link, collected in $\boldsymbol{\zeta}=\{\zeta_{nk}\}$, and gives
\begin{equation}
\label{eq:fq_qt}
\begin{aligned}
f_{\mathrm q}(\mathbf X,\mathbf F,\boldsymbol{\gamma},\boldsymbol{\zeta}) = \sum_{n\in\mathcal N}\sum_{k\in\mathcal K}x_{nk}
\Bigg[
&\log(1+\gamma_{nk})-\gamma_{nk} \\
&+ 2\sqrt{1+\gamma_{nk}}\,\Re\left\{\zeta_{nk}^{\ast}\mathbf h_{nk}^{\mathsf H}\mathbf f_{nk}\right\} \\
&- |\zeta_{nk}|^2 \left(S_{nk}(\mathbf F)+I_{nk}(\mathbf X,\mathbf F)\right)
\Bigg].
\end{aligned}
\end{equation}
For fixed $(\mathbf X,\mathbf F,\boldsymbol{\gamma})$, $f_{\mathrm q}$ is a concave quadratic function of each $\zeta_{nk}$, and its maximizer is
\begin{equation}
\zeta_{nk}^{\star} = \frac{\sqrt{1+\gamma_{nk}}\,\mathbf h_{nk}^{\mathsf H}\mathbf f_{nk}}{S_{nk}(\mathbf F)+I_{nk}(\mathbf X,\mathbf F)}.
\label{eq:zeta_opt}
\end{equation}
Substituting $\zeta_{nk}^{\star}$ back recovers $f_{\mathrm r}$, so maximizing $f_{\mathrm q}$ over $\boldsymbol{\zeta}$ preserves the objective.
The two transforms thus reformulate $(\mathcal P_0)$ equivalently as
\begin{equation}
\begin{aligned}
(\mathcal P_1)\quad \max_{\mathbf X,\mathbf F,\boldsymbol{\gamma},\boldsymbol{\zeta}} \quad &f_{\mathrm q}(\mathbf X,\mathbf F,\boldsymbol{\gamma},\boldsymbol{\zeta}) \label{eq:P1} \\
\mathrm{s.t.}\quad &\eqref{eq:c_assoc}\text{--}\eqref{eq:c_binary}, \\
&\gamma_{nk}\geq0,\quad \zeta_{nk}\in\mathbb C,\quad \forall n\in\mathcal N,\ k\in\mathcal K.
\end{aligned}
\end{equation}

\subsection{Precoder Update}
\label{subsec:precoder_update}
We now solve $(\mathcal P_1)$ by BCD. With $\mathbf{X}$ and $(\boldsymbol{\gamma}, \boldsymbol{\zeta})$ fixed, maximizing $f_{\mathrm q}$ over $\mathbf{F}$ is the convex quadratically constrained quadratic program
\begin{subequations}
\label{eq:P_F}
\begin{align}
\max_{\mathbf{F}} \quad & \sum_{n\in \mathcal{N}} \sum_{k\in \mathcal{K}} x_{nk} \left( 2\Re\left\{ \mathbf{q}_{nk}^{\mathsf{H}}\mathbf{f}_{nk} \right\} - \mathbf{f}_{nk}^{\mathsf{H}}\mathbf{Q}_{nk}\mathbf{f}_{nk} \right) \label{eq:P_F_obj} \\
\mathrm{s.t.} \quad & \sum_{k\in \mathcal{K}} x_{nk}\|\mathbf{f}_{nk}\|^2 \leq P_n, \qquad \forall n\in \mathcal{N}, \label{eq:P_F_power}
\end{align}
\end{subequations}
where $\mathbf{q}_{nk}$ is given by
\begin{equation}
\mathbf{q}_{nk} = \sqrt{1+\gamma_{nk}}\,\zeta_{nk}\mathbf{h}_{nk},
\label{eq:q_nk}
\end{equation}
and $\mathbf{Q}_{nk}$ is given by
\begin{equation}
\mathbf{Q}_{nk} = |\zeta_{nk}|^2 \mathbf{h}_{nk}\mathbf{h}_{nk}^{\mathsf{H}} + \sum_{n'\in \mathcal{N}} \sum_{k'\neq k} x_{n'k'} |\zeta_{n'k'}|^2 \mathbf{h}_{nk'}\mathbf{h}_{nk'}^{\mathsf{H}}.
\label{eq:Q_nk}
\end{equation}
Since $\mathbf{Q}_{nk}\succeq \mathbf{0}$, the objective is concave in $\mathbf{F}$.
The Karush--Kuhn--Tucker (KKT) stationarity condition then gives the optimal precoder of an active link as
\begin{equation}
\mathbf{f}_{nk}^{\star} = \left(\mathbf{Q}_{nk} + \mu_n \mathbf{I}_{M}\right)^{-1} \mathbf{q}_{nk},
\label{eq:f_opt}
\end{equation}
where $\mu_n \geq 0$ is the multiplier of the power constraint of satellite $n$, which can be found by bisection search. The precoder of an inactive link with $x_{nk}=0$ does not affect the transmitted signal, the objective, or the power constraints, and thus does not enter the problem. We therefore keep $\mathbf{f}_{nk}=\mathbf{0}_{M}$ and skip its update.

\subsection{Association Update}
\label{subsec:assoc_update}
We next compute a candidate association for the subsequent precoder update. With $\mathbf{F}$ and $(\boldsymbol{\gamma},\boldsymbol{\zeta})$ fixed, we express $f_{\mathrm q}$ as a function of $\mathbf{X}$. We omit the power constraint~\eqref{eq:c_power} here because the subsequent precoder update enforces it through the multiplier $\mu_n$ under the candidate association. Expanding $S_{nk}$ and $I_{nk}$ reduces the association update to
\begin{subequations}
\label{eq:P_A}
\begin{align}
\max_{\mathbf{X}} \quad & \sum_{n\in\mathcal{N}}\sum_{k\in\mathcal{K}} \xi_{nk}x_{nk} - \sum_{n\in\mathcal{N}}\sum_{k\in\mathcal{K}}\sum_{n'\in\mathcal{N}}\sum_{k'\neq k} \chi_{nk,n'k'}\,x_{nk}x_{n'k'} \label{eq:P_A_obj} \\
\mathrm{s.t.} \quad & \eqref{eq:c_assoc},\ \eqref{eq:c_cap},\ \eqref{eq:c_binary}, \label{eq:P_A_const}
\end{align}
\end{subequations}
where $\xi_{nk}$ is given by
\begin{equation}
\begin{aligned}
\xi_{nk} = {}& \log(1+\gamma_{nk})-\gamma_{nk}+2\sqrt{1+\gamma_{nk}}\,\Re\left\{\zeta_{nk}^{\ast}\mathbf{h}_{nk}^{\mathsf{H}}\mathbf{f}_{nk}\right\} \\
&-|\zeta_{nk}|^2\left(\left|\mathbf h_{nk}^{\mathsf H}\mathbf f_{nk}\right|^2+\sigma^2\right),
\end{aligned}
\label{eq:xi}
\end{equation}
and $\chi_{nk,n'k'}$ is given by
\begin{equation}
\chi_{nk,n'k'} = |\zeta_{nk}|^2\left|\mathbf{h}_{n'k}^{\mathsf{H}}\mathbf{f}_{n'k'}\right|^2.
\label{eq:c_couple}
\end{equation}
Here $\xi_{nk}$ is the gain coefficient of link $(n,k)$ without interference, and $\chi_{nk,n'k'}$ is the loss that the stream of link $(n',k')$ inflicts on it. The objective thus rewards each active link and penalizes each pair of concurrently active links, reflecting their effect on the network sum spectral efficiency. Because these pairwise penalties make the value of a link depend on the other active links, the association remains a binary quadratic optimization problem.

To obtain a tractable association update, we linearize the bilinear term around the current iterate $\mathbf{X}^{(i)}$, with $i$ the iteration index, as
\begin{equation}
x_{nk}x_{n'k'} \approx x_{n'k'}^{(i)}x_{nk} + x_{nk}^{(i)}x_{n'k'} - x_{nk}^{(i)}x_{n'k'}^{(i)},
\label{eq:sca}
\end{equation}
which neglects only the second-order term $(x_{nk}-x_{nk}^{(i)})(x_{n'k'}-x_{n'k'}^{(i)})$. Substituting it into~\eqref{eq:P_A_obj} and dropping the constants yields the linear surrogate
\begin{equation}
\sum_{n\in\mathcal{N}}\sum_{k\in\mathcal{K}} \pi_{nk}^{(i)}x_{nk},
\label{eq:linear_surrogate}
\end{equation}
where the coefficient $\pi_{nk}^{(i)}$ is given by
\begin{equation}
\pi_{nk}^{(i)} = \xi_{nk} - \sum_{n'\in\mathcal{N}}\sum_{k'\neq k}\chi_{nk,n'k'}\,x_{n'k'}^{(i)} - \sum_{n'\in\mathcal{N}}\sum_{k'\neq k}\chi_{n'k',nk}\,x_{n'k'}^{(i)}.
\label{eq:sca_score}
\end{equation}
Here, $\xi_{nk}$ and $\chi_{nk,n'k'}$ are evaluated at the updated precoders, with the corresponding closed-form auxiliary values substituted into the score. The first sum is the interference that user $k$ receives from the other active links, and the second is the interference that link $(n,k)$ causes to the other users. The score $\pi_{nk}^{(i)}$ thus favors a link whose own signal is strong and which neither suffers from nor causes severe interference, and thereby reflects the impact of the link on the entire network.

Maximizing the linear surrogate under the original constraints gives the association update
\begin{equation}
\mathbf{X}^{(i+1)} = \argmax_{\mathbf{X}}\ \sum_{n\in\mathcal{N}}\sum_{k\in\mathcal{K}} \pi_{nk}^{(i)}x_{nk} \quad \mathrm{s.t.}\quad \eqref{eq:c_assoc},\ \eqref{eq:c_cap},\ \eqref{eq:c_binary},
\label{eq:bmatching}
\end{equation}
which is a weighted bipartite b-matching problem between the satellites and the users, i.e., a matching in which each node $v$ is matched at most $b(v)$ times. Each user is matched to at most one satellite, each satellite to at most $K_n$ users, and every activated match $(n,k)$ earns the score $\pi_{nk}^{(i)}$.
The constraint matrix of~\eqref{eq:bmatching} is totally unimodular, and its right-hand side is integral for integer serving caps. Hence, its linear relaxation has binary vertices and admits a binary optimum with the same optimal value, which can be obtained by min-cost flow~\cite{edmonds:jacm:72}.

Premature turning-off occurs when this hard update sets $x_{nk}=0$ and the associated precoder is kept at $\mathbf 0_M$ with its update skipped. The auxiliary updates then give $\gamma_{nk}=\zeta_{nk}=0$, so the gain coefficient $\xi_{nk}$ and both interference terms in~\eqref{eq:sca_score} vanish, leaving $\pi_{nk}^{(i)}=0$.
The subsequent matching therefore assigns no positive gain to activating the link. Since activating a link with zero score leaves the objective in~\eqref{eq:bmatching} unchanged, the solver may still select such a link, i.e., set $x_{nk}=1$. Even in this case, $\zeta_{nk}=0$ gives $\mathbf{q}_{nk}=\mathbf{0}_{M}$ in~\eqref{eq:q_nk}, so the precoder update in~\eqref{eq:f_opt} returns a zero precoder again. Under these block updates, the precoder thus remains zero even if the association variable is subsequently set to one.

Under these updates, a link whose precoder power has dropped to zero cannot recover it, so the links with nonzero power always form a subset of the initial association. A link excluded at initialization thus never receives an optimized precoder, even when changes in the other precoders could make it useful. As a result, the final design is highly sensitive to the initialization. For instance, an initialization based on the signal-to-noise ratio (SNR) groups users by signal strength while ignoring how well they can be separated from one another. To avoid this, the next section keeps the association weights positive while all candidate precoders are updated and defers the binary decision until these candidates have been evaluated.

\section{Soft-then-Hard Association Update}
\label{sec:soft_then_hard}
To mitigate premature turning-off, we retain continuous association weights while the candidate precoders are updated and defer the binary decision to the end of the joint updates. The proposed scheme therefore has two phases. In the soft phase, constraints~\eqref{eq:c_assoc} and~\eqref{eq:c_cap} define a transport problem, which we regularize with entropy and solve by the Sinkhorn algorithm~\cite{peyre:fnt:19,cuturi:2013,sinkhorn-knopp:1967}. Annealing gradually concentrates the weights on links whose scores remain favorable as their precoders are updated. In the hard phase, a single b-matching determines the binary serving links, after which FP refines the selected precoders.

\subsection{Soft Relaxation}
\label{subsec:soft_relaxation}
Relaxing each $x_{nk}$ to an association weight $\tilde{x}_{nk}\in[0,1]$, collected in the soft association $\tilde{\mathbf X}=[\tilde{x}_{nk}]$, turns the b-matching~\eqref{eq:bmatching} into the linear program
\begin{subequations}\label{eq:soft_relax}
\begin{align}
\max_{\tilde{\mathbf X}}\quad & \sum_{n\in\mathcal{N}}\sum_{k\in\mathcal{K}} \pi_{nk}^{(i)}\,\tilde{x}_{nk} \\
\mathrm{s.t.}\quad & \sum_{n\in\mathcal{N}} \tilde{x}_{nk} \le 1, && \forall k\in\mathcal{K}, \label{eq:soft_demand}\\
& \sum_{k\in\mathcal{K}} \tilde{x}_{nk} \le K_n, && \forall n\in\mathcal{N}, \label{eq:soft_capacity}\\
& \tilde{x}_{nk}\in[0,1], && \forall n\in\mathcal{N},\ k\in\mathcal{K}. \label{eq:soft_box}
\end{align}
\end{subequations}
Problem~\eqref{eq:soft_relax} can be viewed as a transport problem, where $\tilde x_{nk}$ is the mass assigned from satellite $n$ to user $k$ and earns score $\pi_{nk}^{(i)}$ per unit mass. Each satellite supplies at most $K_n$ units, and each user receives at most one. These inequalities allow a user to remain unserved and a satellite to use less than its serving cap.
The Sinkhorn algorithm instead requires equality marginals with equal total supply and demand~\cite{peyre:fnt:19}. Simply replacing the inequalities in~\eqref{eq:soft_relax} with equalities would force every user to be served and every satellite to be fully loaded. It would also be infeasible, since the total capacity $\sum_{n\in\mathcal N}K_n$ generally differs from the total demand $K$. We therefore introduce dummy nodes to account for unserved demand and unused capacity while retaining the original bounds on the physical links.

We add a dummy satellite $n=0$ that absorbs the unserved demand and a dummy user $k=0$ that absorbs the unused capacity, and denote the augmented sets by $\tilde{\mathcal N}=\{0\}\cup\mathcal N$ and $\tilde{\mathcal K}=\{0\}\cup\mathcal K$. Specifically, the dummy satellite virtually serves the remaining demand of each user for which~\eqref{eq:soft_demand} is inactive, such as a user selected by no satellite, so that this constraint holds with equality. Likewise, each satellite that does not fill its serving cap, i.e., for which~\eqref{eq:soft_capacity} is inactive, virtually serves the dummy user with its remaining capacity, so that this constraint also holds with equality. The score is extended as $\tilde{\pi}_{nk}=\pi_{nk}^{(i)}$ for the original links $(n,k)\in\mathcal N\times\mathcal K$ and $\tilde{\pi}_{nk}=0$ on the dummy links, since mass routed through a dummy node corresponds to an inactive link. The supply of each satellite and the demand of each user are set to
\begin{equation}
\label{eq:marginals}
\tilde{s}_n=\begin{cases}K_n,& n\in\mathcal N,\\[2pt] K,& n=0,\end{cases}
\qquad
\tilde{d}_k=\begin{cases}1,& k\in\mathcal K,\\[2pt] \sum_{n'\in\mathcal N}K_{n'},& k=0,\end{cases}
\end{equation}
so that the total supply and the total demand both equal $K+\sum_{n\in\mathcal N}K_n$. Denoting the augmented $(N+1)\times(K+1)$ matrix again by $\tilde{\mathbf X}$, we obtain the soft association from the balanced problem
\begin{subequations}
\label{eq:balanced_ot}
\begin{align}
\max_{\tilde{\mathbf X}\ge 0}\quad & \sum_{n\in\tilde{\mathcal N}}\sum_{k\in\tilde{\mathcal K}} \tilde{\pi}_{nk}\,\tilde{x}_{nk}\\
\mathrm{s.t.}\quad & \sum_{k\in\tilde{\mathcal K}} \tilde{x}_{nk}=\tilde{s}_n, && \forall n\in\tilde{\mathcal N},\label{eq:bot_row}\\
& \sum_{n\in\tilde{\mathcal N}} \tilde{x}_{nk}=\tilde{d}_k, && \forall k\in\tilde{\mathcal K}.\label{eq:bot_col}
\end{align}
\end{subequations}
Problem~\eqref{eq:balanced_ot} is a standard optimal transport problem from $\tilde{\mathcal N}$ to $\tilde{\mathcal K}$ with the negative score as the transport cost. Constraint~\eqref{eq:soft_box} is now implied by~\eqref{eq:bot_col} for all users excluding the dummy user, together with $\tilde{\mathbf X}\ge 0$.

\subsection{Entropic Optimal Transport and the Sinkhorn Update}
\label{subsec:sinkhorn}
The balanced linear program still admits a binary optimum on the original links. Its constraint matrix is totally unimodular, and its marginals are integers, so it admits an integral optimal coupling. The unit demands of the original users make their association weights binary. Thus, relaxation alone does not ensure that the candidates retain continuous association weights while their precoders are updated. To retain these weights, we add an entropy term and solve
\begin{equation}
\label{eq:entropic_ot}
\begin{aligned}
\tilde{\mathbf X}^{\star}=\argmax_{\tilde{\mathbf X}\ge 0}\quad & \sum_{n\in\tilde{\mathcal N}}\sum_{k\in\tilde{\mathcal K}}\tilde{\pi}_{nk}\,\tilde{x}_{nk}+\epsilon\,H(\tilde{\mathbf X})\\
\mathrm{s.t.}\quad & \eqref{eq:bot_row},\ \eqref{eq:bot_col},
\end{aligned}
\end{equation}
where $\epsilon>0$ is the regularization parameter and the entropy of the weights is $H(\tilde{\mathbf X})=-\sum_{n,k}\tilde{x}_{nk}(\log\tilde{x}_{nk}-1)$. The regularized problem is strictly concave and has a unique maximizer with positive weights~\cite{cuturi:2013}.
At this positive solution, the nonnegativity constraints are inactive. Associating the multipliers $\lambda_n$ and $\varphi_k$ with the row constraints~\eqref{eq:bot_row} and column constraints~\eqref{eq:bot_col}, respectively, we write the Lagrangian of~\eqref{eq:entropic_ot} as
\begin{equation}
\label{eq:ot_lagrangian}
\begin{aligned}
\mathcal L ={}& \sum_{n\in\tilde{\mathcal N}}\sum_{k\in\tilde{\mathcal K}}\tilde{\pi}_{nk}\,\tilde x_{nk}+\epsilon\,H(\tilde{\mathbf X})\\
&-\sum_{n\in\tilde{\mathcal N}}\lambda_n\Big(\sum_{k\in\tilde{\mathcal K}}\tilde x_{nk}-\tilde s_n\Big)
-\sum_{k\in\tilde{\mathcal K}}\varphi_k\Big(\sum_{n\in\tilde{\mathcal N}}\tilde x_{nk}-\tilde d_k\Big).
\end{aligned}
\end{equation}
Since $\frac{\partial H(\tilde{\mathbf X})}{\partial\tilde x_{nk}}=-\log\tilde x_{nk}$, setting $\frac{\partial\mathcal L}{\partial\tilde x_{nk}}=0$ gives the stationarity condition
\begin{equation}
\label{eq:ot_stationarity}
\tilde{\pi}_{nk}-\epsilon\log\tilde{x}_{nk}-\lambda_n-\varphi_k=0,
\end{equation}
and solving it for $\tilde x_{nk}$ gives
\begin{equation}
\label{eq:gibbs}
\tilde{x}_{nk}^{\star}=\exp\!\Big(\frac{\tilde{\pi}_{nk}-\lambda_n-\varphi_k}{\epsilon}\Big)=u_n\,\Phi_{nk}\,v_k,
\end{equation}
where $\Phi_{nk}=\exp(\tilde{\pi}_{nk}/\epsilon)$, and the positive scaling factors $u_n=\exp(-\lambda_n/\epsilon)$ and $v_k=\exp(-\varphi_k/\epsilon)$ are collected in $\mathbf u\in\mathbb R_{+}^{N+1}$ and $\mathbf v\in\mathbb R_{+}^{K+1}$. Each weight is thus the exponential of its score, scaled by a factor $u_n$ of satellite $n$ and a factor $v_k$ of user $k$, and it remains to choose these factors so that the marginal constraints hold. Multiplying $\mathbf u$ by any positive constant and dividing $\mathbf v$ by the same constant leaves every weight unchanged, so the scaling factors are determined only up to such a constant, whereas the solution itself is unique. Substituting~\eqref{eq:gibbs} into~\eqref{eq:bot_row} and~\eqref{eq:bot_col} gives
\begin{equation}
\label{eq:sinkhorn}
u_n=\frac{\tilde{s}_n}{\sum_{k\in\tilde{\mathcal K}}\Phi_{nk}v_k},\qquad
v_k=\frac{\tilde{d}_k}{\sum_{n\in\tilde{\mathcal N}}\Phi_{nk}u_n}.
\end{equation}
Each equation gives one scaling vector in closed form when the other is fixed, so the Sinkhorn algorithm~\cite{sinkhorn-knopp:1967} solves~\eqref{eq:sinkhorn} by updating $\mathbf u$ and $\mathbf v$ alternately. The iterates converge to the scaling factors that satisfy both equations~\cite{franklin-lorenz:1989}, and substituting them into~\eqref{eq:gibbs} gives the solution of~\eqref{eq:entropic_ot}.

Three properties of the soft association update help prevent premature turning-off.
First, $\Phi_{nk}>0$ ensures that every association weight remains strictly positive, so the precoder and auxiliary variables of every candidate link continue to be updated.
Second, a small association weight does not directly scale down the corresponding precoder. In the precoder problem~\eqref{eq:P_F}, $\tilde{x}_{nk}$ multiplies both the objective term and the power term of link $(n,k)$, so it cancels out in the stationarity condition. As a result, the closed-form precoder~\eqref{eq:f_opt} does not contain $\tilde{x}_{nk}$ as a direct multiplier. Instead, the weights enter only through the interference terms in $\mathbf{Q}_{nk}$ and the shared power budget. Every candidate link is therefore evaluated using its own optimized precoder, which prevents links with small weights from losing power and dying out.
Third, $\epsilon$ controls how evenly the weights are distributed. A large $\epsilon$ keeps candidate links actively competing while their precoders are updated, whereas a small $\epsilon$ drives the soft association toward hard matching~\cite{peyre:fnt:19}. Gradually decreasing $\epsilon$ delays hard decisions until the precoders have been sufficiently optimized.

\subsection{Annealing}
\label{subsec:annealing}
To benefit from both a large and a small $\epsilon$, we start from a large initial value $\epsilon^{(0)}$ and decrease $\epsilon$ over the outer iterations of the soft phase as
\begin{equation}
\epsilon^{(i)} = \max\!\big(\epsilon^{(i-1)}/\rho,\ \epsilon_{\min}\big),\qquad \rho>1,
\label{eq:anneal}
\end{equation}
where $\rho$ is the annealing factor and $\epsilon_{\min}$ is the floor.

\begin{remark}[Role of annealing] \normalfont
With the scores fixed, the Sinkhorn algorithm solves~\eqref{eq:entropic_ot} for every $\epsilon>0$, so annealing is not required for this inner problem. Instead, annealing keeps the concentration of the association in step with the precoder updates as the scores evolve over the outer iterations.
In early iterations, the scores rely on precoders that are not yet optimized. If $\epsilon$ is small from the start, a self-reinforcing loop may arise: the association concentrates on the links favored by the initial scores, the precoders are optimized for this association, and the resulting scores confirm the same links. A large initial $\epsilon$ instead spreads the weights over the candidates so that their precoders can be optimized before the association concentrates. As $\epsilon$ is gradually reduced, the preference moves toward links that still score high after this optimization, in line with deterministic annealing~\cite{rose:procieee:98}. Unlike per-iteration hard matching, this procedure evaluates the candidates before the final binary selection is made.
\end{remark}

\subsection{Hard Mapping and Overall Algorithm}
\label{subsec:hard_algo}
The soft phase alternates the auxiliary, precoder, and annealed association updates, using $\tilde{\mathbf X}$ in every block, until $\epsilon$ reaches its floor and the soft association converges. These weights support candidate evaluation, but the final serving links must satisfy the binary constraint~\eqref{eq:c_binary}. Hard mapping then converts the evaluated candidates into a feasible binary association.
At the final soft association $\tilde{\mathbf X}^{(i)}$, we update the auxiliary variables and precoders and evaluate the scores $\pi_{nk}^{\star}$ at this operating point. Setting $\epsilon=0$ removes the entropy term from the association subproblem. The resulting balanced linear program admits an integral optimum, whose entries on the original links give the binary association through the b-matching
\begin{equation}
\label{eq:hard_map}
\mathbf X^{\star}=\argmax_{\mathbf X}\ \sum_{n\in\mathcal N}\sum_{k\in\mathcal K}\pi_{nk}^{\star}\,x_{nk}\quad\mathrm{s.t.}\ \eqref{eq:c_assoc},\ \eqref{eq:c_cap},\ \eqref{eq:c_binary}.
\end{equation}
The matching makes a single binary selection using scores obtained after the joint soft updates, rather than repeatedly fixing the serving links while the candidate precoders are still being updated. Moving from continuous association weights to binary links can change both the interference pattern and the allocation of each satellite's power budget. With $\mathbf X^{\star}$ fixed, FP refines the selected precoders for this binary association and gives $\mathbf F^{\star}$. The auxiliary and precoder updates do not decrease the objective for the fixed association~\cite{shen:tsp:18:part1}. Algorithm~\ref{alg:soft_then_hard} summarizes the procedure.

\begin{algorithm}[t]
\DontPrintSemicolon
\SetArgSty{textnormal}
\caption{Soft-then-Hard Joint Association and Precoding}
\label{alg:soft_then_hard}
\KwIn{channels $\{\mathbf h_{nk}\}$, power budgets $\{P_n\}$, serving caps $\{K_n\}$, annealing parameters $\epsilon^{(0)}$, $\rho$, $\epsilon_{\min}$}
\KwOut{association $\mathbf X^{\star}$, precoders $\mathbf F^{\star}$}
Initialize $\tilde{\mathbf X}$ and $\mathbf F$, set $\tilde{\mathbf s}$ and $\tilde{\mathbf d}$ by \eqref{eq:marginals}, and set $i=0$\;
\Repeat{$\epsilon^{(i)}=\epsilon_{\min}$ and $\tilde{\mathbf X}$ converges}{
Update $\boldsymbol\gamma$ and $\boldsymbol\zeta$ by \eqref{eq:gamma_opt} and \eqref{eq:zeta_opt}\;
Update $\mathbf F$ by \eqref{eq:f_opt}\;
Update the scores $\tilde\pi_{nk}$ by \eqref{eq:sca_score}\;
Update $\epsilon^{(i+1)}$ by \eqref{eq:anneal}\;
Update $\tilde{\mathbf X}^{(i+1)}$ by \eqref{eq:gibbs} and \eqref{eq:sinkhorn}, and set $i\leftarrow i+1$\;
}
Update $\boldsymbol\gamma$, $\boldsymbol\zeta$, $\mathbf F$, and the scores $\pi_{nk}^{\star}$ under $\tilde{\mathbf X}^{(i)}$\;
Obtain $\mathbf X^{\star}$ by \eqref{eq:hard_map}\;
Update $\boldsymbol\gamma$, $\boldsymbol\zeta$, and $\mathbf F$ with $\mathbf X=\mathbf X^{\star}$ until convergence, which gives $\mathbf F^{\star}$\;
\Return $\mathbf X^{\star}$, $\mathbf F^{\star}$
\end{algorithm}

\subsection{Complexity Analysis}
\label{subsec:complexity}
In the precoder update~\eqref{eq:f_opt}, solving the $K \times K$ system of equations for the combining coefficients requires $\mathcal{O}(K^3)$ operations per satellite, after which constructing the $K$ precoders costs $\mathcal{O}(MK^2)$. Computing the precoder inner products and auxiliary variables across all satellites adds $\mathcal{O}(NMK^2)$. The precoder update thus incurs a total complexity of $\mathcal{O}(N(K^3 + MK^2))$ per outer iteration, which can be further accelerated using recent fast FP methods~\cite{shen:jsac:24:accelerated, chen:twc:25:fastfp}.

Regarding link association, the Sinkhorn association update requires $\mathcal{O}(NK)$ operations per Sinkhorn iteration via row and column scaling. Although the b-matching can be solved in polynomial time by min-cost flow, the proposed scheme executes this matching only once after the soft phase. Because this one-time operation contributes a negligible cost over the entire execution, the overall asymptotic complexity of the proposed scheme is governed by the precoder updates, remaining $\mathcal{O}(N(K^3 + MK^2))$. In contrast, hard matching with FP precoding solves this min-cost flow problem at every outer iteration.

Finally, we consider greedy link addition for comparison. If the sequential link addition in~\cite{zhang:twc:24:multisat} is combined with our FP solver, up to $NK$ candidate links must be tested in each of at most $K$ stages. Since the precoders are recomputed for every tested link, this approach has a complexity of $\mathcal{O}(N^2K^4(M+K))$. The proposed soft-then-hard scheme updates all candidate links at once and thus avoids such repeated precoder computations.

\section{Space-Time Precoding for LEO Links}
\label{sec:stab}
As noted in Remark~\ref{rem:cap}, the users of a cell lie in nearly the same direction from the satellite, so spatial precoding alone separates them poorly even within the serving cap~\cite{you:jsac:20}. Their Doppler shifts can nevertheless differ because of differences in the user velocities and LoS directions, providing an additional means of separating the users through the temporal dimension. We exploit this through space-time beamforming~\cite{yim:twc:26,yun:twc:26}, which repeats each symbol over multiple snapshots so that the Doppler shift of each link appears as a distinct phase progression. We obtain the Doppler shift of each link by projecting the satellite--user relative velocity onto the link direction, thereby also capturing the Doppler differences caused by the satellite motion. We first stack these snapshots into a space-time channel and then integrate it into the proposed joint design, where the same FP and soft-then-hard updates jointly optimize association and space-time precoding.

\subsection{LEO Space-Time Channel Construction} \label{subsec:channel_model}
We transmit each symbol $L$ times at the instants $t_\ell=(\ell-1)T_{\mathrm r}$, $\ell=1,\dots,L$, where $T_{\mathrm r}$ is the repetition interval. Over the window $(L-1)T_{\mathrm r}$, we assume that the composite gain remains equal to its reference value $\beta_{nk}=\beta_{nk}(0)$ and that $\nu_{nk}$ and $\mathbf a_{nk}$ remain constant. From~\eqref{eq:reduced}, the snapshot channel $\mathbf g_{nk}[\ell]=\mathbf g_{nk}(t_\ell)$ is then
\begin{equation}
\mathbf{g}_{nk}[\ell] = \sqrt{M}\,\beta_{nk}\,e^{j2\pi(\ell-1)\omega_{nk}}\,
\mathbf{a}_{nk},\quad \ell=1,\dots,L,
\label{eq:snapshot}
\end{equation}
where $\omega_{nk}=\nu_{nk}T_{\mathrm r}$ is the normalized Doppler shift. Collecting the snapshot phases into the temporal response vector
\begin{equation}
\mathbf{b}_{nk} = \frac{1}{\sqrt{L}}
\big[1,\, e^{j2\pi\omega_{nk}},\, \dots,\, e^{j2\pi(L-1)\omega_{nk}}\big]^{\mathsf{T}}
\in\mathbb{C}^{L}
\label{eq:temporal}
\end{equation}
and stacking the $L$ snapshot channels gives the space-time channel
\begin{align}
\bar{\mathbf{h}}_{nk}
&= \big[\mathbf{g}_{nk}[1]^\mathsf{T},\, \dots,\, \mathbf{g}_{nk}[L]^\mathsf{T}\big]^\mathsf{T}
\notag\\
&= \sqrt{ML}\,\beta_{nk}\,\big(\mathbf{b}_{nk}\otimes\mathbf{a}_{nk}\big)
\in\mathbb{C}^{ML}.
\label{eq:st_channel}
\end{align}
The Kronecker structure combines the spatial response $\mathbf a_{nk}$ with the temporal response $\mathbf b_{nk}$ determined by the Doppler shift.

\subsection{Integration with the Proposed Joint Design}
\label{subsec:stab_integration}
The space-time channel $\bar{\mathbf h}_{nk}\in\mathbb C^{ML}$ replaces the spatial channel $\mathbf h_{nk}\in\mathbb C^{M}$, and the rest of the signal model follows. At the $\ell$-th snapshot, satellite $n$ applies the precoder $\mathbf{f}_{nk}[\ell]\in\mathbb{C}^{M}$ to the symbol of user $k$, and the precoders of the $L$ snapshots form the space-time precoder
\begin{equation}
\bar{\mathbf{f}}_{nk} = \frac{1}{\sqrt{L}}\big[\mathbf{f}_{nk}[1]^{\mathsf{T}},\dots,
\mathbf{f}_{nk}[L]^{\mathsf{T}}\big]^{\mathsf{T}}\in\mathbb{C}^{ML},
\label{eq:precoder}
\end{equation}
whose squared norm is the transmit power of the link averaged over the snapshots, so the power constraint~\eqref{eq:c_power} limits the average transmit power of each satellite. At the $\ell$-th snapshot, user $k$ receives
\begin{equation}
y_k[\ell] = \sum_{n\in\mathcal{N}}\sum_{k'\in\mathcal{K}} x_{nk'}\,
\mathbf{g}_{nk}[\ell]^{\mathsf{H}}\mathbf{f}_{nk'}[\ell]\, s_{k'} + z_k[\ell],
\label{eq:rx_snapshot}
\end{equation}
where $z_k[\ell]\sim\mathcal{CN}(0,\sigma^2)$ is independent across snapshots. User $k$ combines the $L$ snapshots coherently as
\begin{equation}
\bar{y}_k=\frac{1}{\sqrt{L}}\sum_{\ell=1}^{L}y_k[\ell]
=\sum_{n\in\mathcal{N}}\sum_{k'\in\mathcal{K}} x_{nk'}\,
\bar{\mathbf{h}}_{nk}^{\mathsf{H}}\bar{\mathbf{f}}_{nk'}\,s_{k'}+\bar{z}_k,
\label{eq:combined}
\end{equation}
where the combined noise $\bar{z}_k$ has variance $\sigma^2$ because the combining weights have unit norm. The combined observation has the form of the spatial received signal~\eqref{eq:rx} with $\bar{\mathbf h}_{nk}$ and $\bar{\mathbf f}_{nk}$ in place of $\mathbf h_{nk}$ and $\mathbf f_{nk}$, and we collect the space-time precoders in $\bar{\mathbf F}$.

Since each symbol occupies $L$ channel uses, the sum spectral efficiency of the space-time scheme is
\begin{equation}
\bar{R}_{\Sigma}(\mathbf{X},\bar{\mathbf{F}}) = \frac{1}{L}
\sum_{n\in\mathcal{N}}\sum_{k\in\mathcal{K}} x_{nk}\,
\log\!\big(1+\overline{\mathrm{SINR}}_{nk}(\mathbf{X},\bar{\mathbf{F}})\big).
\label{eq:sumse_st}
\end{equation}
Here, $\overline{\mathrm{SINR}}_{nk}$ is given by~\eqref{eq:sinr} with $\bar{\mathbf h}_{nk}$ and $\bar{\mathbf f}_{nk}$ in place of $\mathbf h_{nk}$ and $\mathbf f_{nk}$. The FP updates retain their form with precoders of dimension $ML$. The association scores in~\eqref{eq:sca_score} are multiplied by $1/L$, so~\eqref{eq:entropic_ot} uses the rate scale of~\eqref{eq:sumse_st}.

The additional separability follows from the Kronecker structure in~\eqref{eq:st_channel}. For two links $(n,k)$ and $(n,k')$ of the same satellite, the mixed-product property gives the normalized space-time channel correlation
\begin{equation}
\frac{|\bar{\mathbf h}_{nk}^{\mathsf H}\bar{\mathbf h}_{nk'}|}{\|\bar{\mathbf h}_{nk}\|\,\|\bar{\mathbf h}_{nk'}\|}
= |\mathbf a_{nk}^{\mathsf H}\mathbf a_{nk'}|
\left|\frac{\sin(\pi L\Delta\omega_{n,kk'})}{L\sin(\pi\Delta\omega_{n,kk'})}\right|,
\label{eq:st_corr}
\end{equation}
where $\Delta\omega_{n,kk'}=\omega_{nk}-\omega_{nk'}$. The first factor is the spatial correlation, which is strong for users in the same cell. The second factor is the temporal correlation, a Dirichlet kernel of the Doppler difference that vanishes when $\Delta\omega_{n,kk'}$ is a multiple of $1/L$. Hence, even when the spatial correlation is strong, such users become nearly orthogonal in the space-time domain, and the precoder can suppress their mutual interference without sacrificing much of the desired signal power. A larger $L$ resolves smaller Doppler differences but reduces the pre-log factor $1/L$ in~\eqref{eq:sumse_st}.

This separability, however, depends on which users are grouped together. The correlation in~\eqref{eq:st_corr} is a pairwise property, and both $\mathbf a_{nk}$ and $\omega_{nk}$ depend on the serving satellite, since $\nu_{nk}$ is determined by the relative velocity between satellite $n$ and user $k$ along their LoS direction. The same pair of users can therefore be well separated by one satellite and poorly separated by another. An association that selects each link by its own channel gain overlooks these pairwise Doppler differences, and precoders optimized for such an association cannot recover them. The Doppler separability must therefore be accounted for when the association is decided, not only when the precoders are designed. In the proposed design, the scores in~\eqref{eq:sca_score} are evaluated with the space-time precoders and account for the interference that each link receives and causes, so the association update favors user groups that are separable in the space-time domain. The $1/L$ scaling keeps the scores on the rate scale of~\eqref{eq:sumse_st}, so the same annealing schedule applies for any $L$.

\section{Numerical Results}
\label{sec:results}

\begin{figure}[!t]
\centering
\includegraphics[width=\columnwidth]{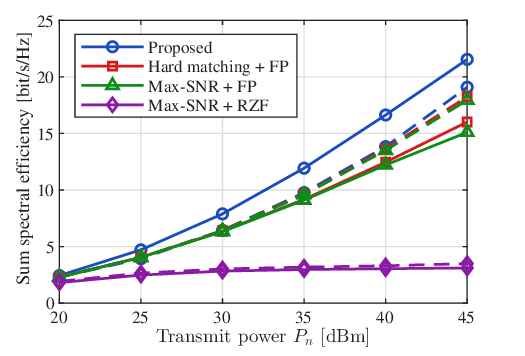}
\caption{Average sum spectral efficiency versus the transmit power $P_n$. Solid lines show $L=3$ and dashed lines show $L=1$.}
\label{fig:power}
\end{figure}
\begin{figure}[!t]
\centering
\includegraphics[width=\columnwidth]{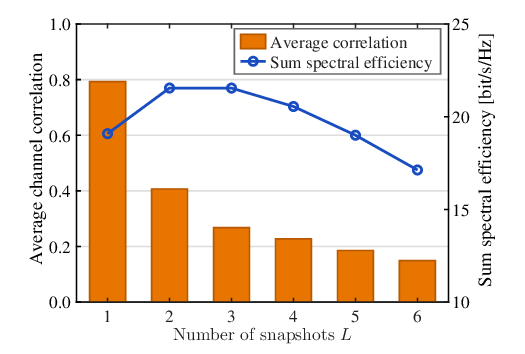}
\caption{Average channel correlation and sum spectral efficiency of the proposed scheme versus $L$.}
\label{fig:doppler_groups}
\end{figure}
\begin{figure}[!t]
\centering
\includegraphics[width=\columnwidth]{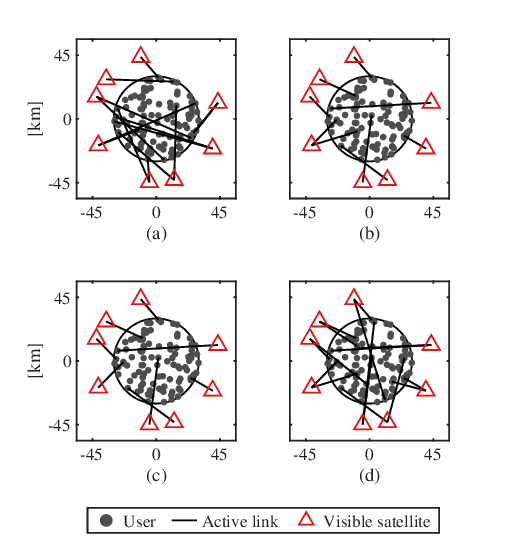}
\caption{Final link association of one realization for (a) the proposed scheme, (b) hard matching with FP precoding, (c) max-SNR association with FP precoding, and (d) max-SNR association with RZF precoding. A link is shown as active if its user attains a spectral efficiency of at least $0.1$~bit/s/Hz.}
\label{fig:assoc}
\end{figure}
\begin{figure*}[!t]
\centering
\includegraphics[width=\textwidth]{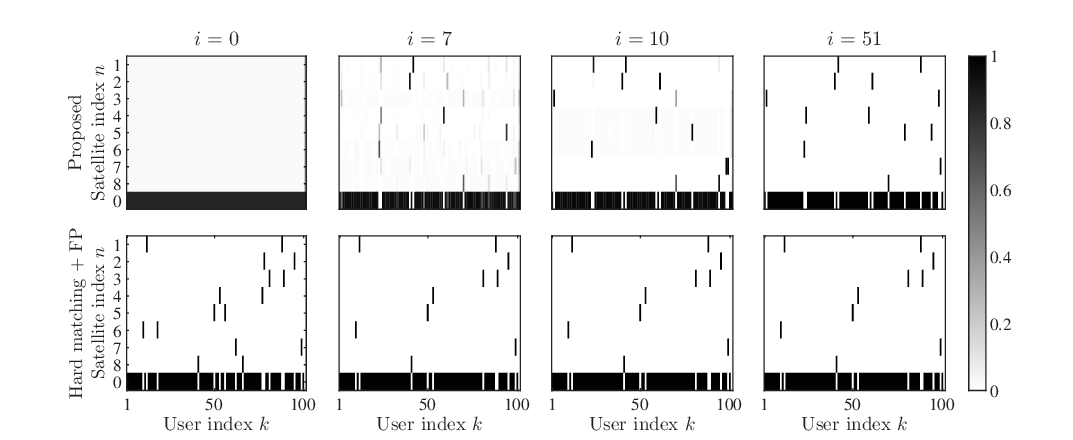}
\caption{Evolution of the association for one realization under the proposed scheme (top) and hard matching with FP precoding (bottom). The last column shows the proposed association after hard mapping. Rows and columns represent satellites and users, respectively, with the dummy nodes in the last row and column. Association weights range from $0$ (white) to $1$ (black).}
\label{fig:heatmap}
\end{figure*}
\begin{figure}[!t]
\centering
\includegraphics[width=\columnwidth]{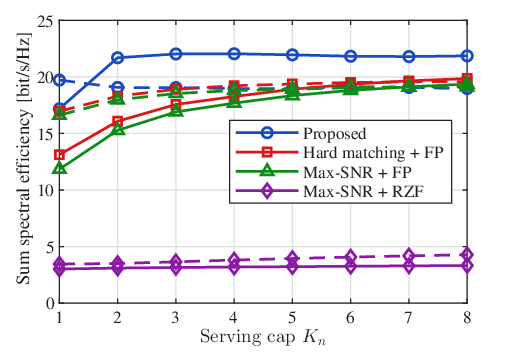}
\caption{Average sum spectral efficiency versus the serving cap $K_n$. Solid lines show $L=3$ and dashed lines show $L=1$.}
\label{fig:mnsweep}
\end{figure}
\begin{figure}[!t]
\centering
\includegraphics[width=\columnwidth]{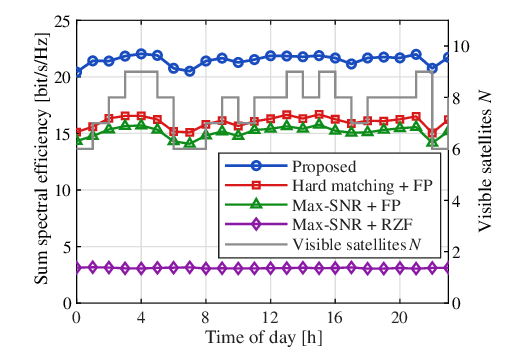}
\caption{Average sum spectral efficiency versus the time of day with $L=3$, together with the number of visible satellites $N$.}
\label{fig:tod}
\end{figure}
\begin{figure}[!t]
\centering
\includegraphics[width=\columnwidth]{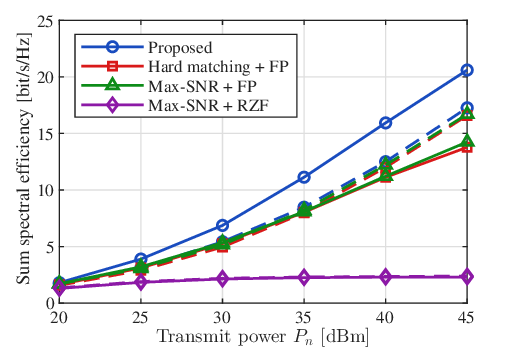}
\caption{Average sum spectral efficiency versus the transmit power $P_n$ under the two-timescale operation with $K_n=2$. Solid lines show $L=3$ and dashed lines show $L=1$.}
\label{fig:twoscale}
\end{figure}
\subsection{Simulation Setting}
\label{subsec:sim_setting}
We consider a Ka-band downlink at a carrier frequency of $20$~GHz from a single-shell Walker Delta constellation $53^\circ{:}3360/28/1$ that follows the SpaceX Gen2 filing~\cite{walker:jbis:84, fcc:spacex-gen2-order:2022}. The shell consists of $3360$ satellites in $28$ evenly spaced orbital planes, with $120$ satellites per plane and a phasing parameter of $1$, at an altitude of $525$~km and an inclination of $53^\circ$. The constellation serves earth stations in motion (ESIM) in a circular cell centered at Seoul ($37.57^\circ$N, $126.98^\circ$E), and the satellites whose elevation angle at the cell center is at least $\vartheta_{\min}=40^\circ$ are visible. The users are dropped uniformly in the cell and belong to three ESIM classes, namely vehicular ($0$--$44$~m/s), maritime ($0$--$20$~m/s), and aircraft ($200$--$280$~m/s at $9$--$12$~km altitude), with probabilities $0.5$, $0.3$, and $0.2$.

The bandwidth is $400$~MHz, and the noise power density is $-174$~dBm/Hz. Each satellite carries a $16\times16$ UPA ($M=256$), the user antenna gain is $G_{\mathrm r}=36.8$~dBi, and the Rician factor is $\kappa_{nk}=15$~dB. We set $R=30$~km, $K=100$, $K_n=2$, and $P_n=45$~dBm. Space-time precoding uses $L=3$ snapshots separated by $T_{\mathrm r}=\SI{100}{\micro\second}$, while $L=1$ corresponds to spatial precoding only. The annealing parameters are set to $\epsilon^{(0)}=1$, $\rho=1.5$, and $\epsilon_{\min}=10^{-3}$. At initialization, the proposed scheme assigns every candidate link a positive association weight that is independent of the channel strength and satisfies the transport constraints, and sets the precoders to maximum ratio transmission (MRT) scaled to meet the weighted power budget of each satellite. The sum spectral efficiency is reported in bit/s/Hz, i.e., the objective divided by $\ln2$, and averaged over channel realizations.

We compare the proposed design with the following three baselines.
\begin{itemize}
\item \textbf{Hard matching with FP precoding}: This baseline starts from a max-SNR association and uses the same auxiliary and precoder updates, but applies the b-matching at every iteration. It isolates the effect of the soft-then-hard association update.
\item \textbf{Max-SNR association with FP precoding}: This baseline selects the links that maximize the sum of their channel gains under constraints~\eqref{eq:c_assoc} and~\eqref{eq:c_cap}, and then optimizes the precoders for this fixed association. It reveals the benefit of jointly optimizing association and precoding.
\item \textbf{Max-SNR association with regularized zero-forcing (RZF) precoding}: This baseline applies RZF precoding to the same fixed association. It reveals the benefit of the FP precoder.
\end{itemize}

\subsection{Performance Evaluation}
\label{subsec:results}

\fig{fig:power} compares the average sum spectral efficiency as the transmit power increases. The proposed scheme with $L=3$ achieves the highest performance throughout the considered range, and its advantage becomes more pronounced at high power, where interference suppression is increasingly important. The RZF baseline performs much worse since each satellite suppresses interference only among its own users, while inter-satellite interference is significant under FFR.

The comparison between $L=1$ and $L=3$ shows that the space-time gain depends on the association. Increasing $L$ from one to three improves the proposed scheme at every power, whereas both FP baselines perform worse with $L=3$ than with $L=1$ at high power. Space-time precoding is beneficial only when the SINR improvement from Doppler separation outweighs the repetition cost $1/L$, and this improvement depends on which users are grouped on each satellite. Max-SNR association groups users by their individual SNRs regardless of their Doppler differences, and hard matching cannot add links beyond this initial association because of premature turning-off. Their user groups therefore gain too little from the temporal dimension to offset the repetition cost at high power. In contrast, the proposed scheme evaluates candidate links with their space-time precoders before hard mapping, so it forms user groups that can be separated through their Doppler shifts.

\fig{fig:doppler_groups} shows how the separability of the selected user groups and the sum spectral efficiency vary with $L$, where association and precoding are jointly optimized for each $L$. The correlation is the magnitude of the normalized inner product between the channel vectors of the two users associated with a satellite, averaged with equal weight over all satellites serving two users.

As $L$ increases, the mean correlation decreases, meaning that the Doppler phase progressions make the selected users easier to distinguish. The sum spectral efficiency, however, peaks at $L=2,3$ and declines afterward even though the correlation keeps decreasing. For larger $L$, the additional gain from better separation can no longer offset the loss due to the pre-log factor $1/L$. Hence, the number of snapshots $L$ should be chosen by trading off Doppler separation against the repetition cost.

\fig{fig:assoc} illustrates the resulting associations for one realization. The proposed scheme serves two users on most satellites, whereas the FP baselines mostly serve one user per satellite. This shows that the proposed scheme better selects users that can be served together, thereby serving more users simultaneously than the FP baselines. The RZF baseline also activates two links on most satellites, but their spectral efficiencies remain low because inter-satellite interference is not suppressed.

\fig{fig:heatmap} traces the association of one realization over the outer iterations. Starting from diffuse weights, the proposed scheme progressively concentrates on a smaller set of links as the regularization parameter decreases. The intermediate patterns still change before the final hard mapping, allowing the association to be reconsidered as the precoders are updated. In contrast, the pattern of hard matching stabilizes early, and the discarded links do not return.

\fig{fig:mnsweep} examines the effect of the serving cap $K_n$. With $L=1$, increasing $K_n$ yields only marginal gains, and all schemes saturate. As noted in Remark~\ref{rem:cap}, the users of a cell are strongly correlated in the spatial domain, so the additional users admitted by a larger cap mainly add interference that spatial precoding cannot suppress. With $L=3$, the proposed scheme overcomes this limitation by serving users that are separated through their Doppler shifts. Its sum spectral efficiency rises sharply as $K_n$ increases from one to two and then stays above the saturation level of spatial precoding. It also outperforms the baselines at every cap, with a gain of $34.9\%$ over hard matching at $K_n=2$.

\fig{fig:tod} evaluates the schemes over a day under changing satellite visibility. The proposed scheme maintains its advantage throughout the day, while its performance and that of hard matching broadly follow the satellite visibility. Additional satellites provide both candidate links and transmit power, but also introduce interference. The sustained gain shows that the proposed association update remains effective as the constellation geometry changes.

Finally, \fig{fig:twoscale} evaluates a two-timescale operation, in which the association is updated on the slow timescale of the link geometry and the precoders on the fast timescale of fading. On the slow timescale, the proposed scheme determines the association by performing the soft phase and hard mapping on reference channels built from the array responses, Doppler shifts, and average channel powers. On the fast timescale, this association is held fixed while FP updates the precoders with instantaneous CSI. The sum spectral efficiency is averaged over fading realizations and geometries. Compared with \fig{fig:power}, where the association is updated for every channel realization, the proposed scheme with $L=3$ achieves nearly the same sum spectral efficiency at high transmit power and retains its advantage over the baselines. The user combinations selected on the slow timescale thus remain effective under fading, so instantaneous channel variations can be handled by the precoder updates alone without updating the association.

\FloatBarrier
\section{Conclusion}
\label{sec:conclusion}
We have presented a joint association and precoding framework for multi-satellite LEO downlinks under FFR. Its key idea is to keep the candidate precoders while the association is updated, so that links are not irreversibly lost by zeroing their precoders after early hard decisions. Entropy regularization and annealing determine when the association concentrates, and a final matching imposes the binary serving constraints. This mechanism and space-time precoding complement each other: Doppler differences add separation among spatially clustered users, and the soft association enables the algorithm to find user combinations that exploit this separation. At the nominal serving cap, the resulting SINR gains outweigh the repetition cost and lead to a higher sum spectral efficiency than the considered baselines. As the limited onboard power makes energy efficiency a key concern in satellite communications, extending the framework to energy efficiency is an important direction for future work.

\bibliographystyle{IEEEtran}
\bibliography{ref_leo_link_asso_precoding}

\end{document}